\documentclass[10pt, conference]{IEEEtran}
\usepackage{graphicx,subfig}
\usepackage{cite}
\usepackage{color}
\usepackage{bigstrut}
\usepackage{multirow}
\usepackage{amsmath}
\usepackage{url}
\usepackage{array}
\usepackage[linesnumbered,ruled,vlined]{algorithm2e}
\SetKwInput{KwInput}{Input}
\SetKwInput{KwOutput}{Output} 
\newcolumntype{L}[1]{>{\raggedright\let\newline\\\arraybackslash\hspace{0pt}}m{#1}}
\newcolumntype{C}[1]{>{\centering\let\newline\\\arraybackslash\hspace{0pt}}m{#1}}
\newcolumntype{R}[1]{>{\raggedleft\let\newline\\\arraybackslash\hspace{0pt}}m{#1}}
\usepackage{amssymb}
\DeclareMathOperator*{\argmax}{argmax}
\DeclareMathOperator*{\argmin}{argmin}

\begin{document}

\title{White-Box Adversarial Attacks on Deep Learning-Based Radio Frequency Fingerprint Identification}

\author{
\IEEEauthorblockN{
Jie~Ma\IEEEauthorrefmark{1},
Junqing~Zhang\IEEEauthorrefmark{1},
Guanxiong~Shen\IEEEauthorrefmark{1},
Alan~Marshall\IEEEauthorrefmark{1}, and
Chip-Hong~Chang\IEEEauthorrefmark{2}
}

\IEEEauthorblockA{
\IEEEauthorrefmark{1}
Department of Electrical Engineering and Electronics, University of Liverpool, Liverpool, L69 3GJ, United Kingdom\\ 
Email: \{Jie.Ma, Junqing.Zhang, Guanxiong.Shen, Alan.Marshall\}@liverpool.ac.uk}
\IEEEauthorrefmark{2}
School of Electrical and Electronic Engineering, Nanyang Technological University, Singapore\\ 
Email: echchang@ntu.edu.sg}

\maketitle

\begin{abstract}
Radio frequency fingerprint identification (RFFI) is an emerging technique for the lightweight authentication of wireless Internet of things (IoT) devices. RFFI exploits unique hardware impairments as device identifiers, and deep learning is widely deployed as the feature extractor and classifier for RFFI.
However, deep learning is vulnerable to adversarial attacks, where adversarial examples are generated by adding perturbation to clean data for causing the classifier to make wrong predictions. 
Deep learning-based RFFI has been shown to be vulnerable to such attacks, however, there is currently no exploration of effective adversarial attacks against a diversity of RFFI classifiers.
In this paper, we report on investigations into white-box attacks (non-targeted and targeted) using two approaches, namely the fast gradient sign method (FGSM) and projected gradient descent (PGD). A LoRa testbed was built and real datasets were collected. These adversarial examples have been experimentally demonstrated to be effective against convolutional neural networks (CNNs), long short-term memory (LSTM) networks, and gated recurrent units (GRU).
\end{abstract}
	

\section{Introduction}
\IEEEPARstart{I}nternet of things (IoT) devices are deployed widely in smart homes and cities, connected healthcare, and industry 4.0, etc.~\cite{baker2017internet}. IoT networks may carry private, sensitive and/or confidential information, and device authentication is the first line of defence against spoofing attacks in an IoT network. Conventional authentication methods rely on cryptographic algorithms and software addresses such as media access control (MAC) addresses. However, cryptographic algorithms require a secure key distribution mechanism, which may not be affordable or practical for low-cost IoT devices~\cite{zhang2020new}.
Software addresses can be easily tampered with. Thus, there is a strong need for a lightweight IoT authentication scheme~\cite{hassija2019survey}. 


Radio frequency fingerprint identification (RFFI) is an emerging device authentication technique~\cite{9450821}. 
The transmitter chain of a wireless device consists of numerous analogue components such as mixers, oscillators, power amplifiers, etc. These components deviate slightly from their nominal specifications due to their inevitable manufacturing processes variations, such as IQ imbalance, carrier frequency offset (CFO), and PA non-linearity~\cite{9450821}. An RFFI system can extract and analyze the unique distortion induced by these imperfections onto the emitted waveforms to identify different devices~\cite{9450821}.
The RFFI process is deployed at the receiver and requires no modification to the IoT end nodes. Machine learning based feature extractors have been used for RFFI as they are communication protocols agnostic and require no domain knowledge to train. Deep learning techniques~\cite{jian2020deep,shen2021radio,roy2019rf}, such as convolutional neural networks (CNNs)~\cite{shen2021radio,shen2021jsac}, long-short term memory (LSTM) networks~\cite{shen2021jsac}, gated recurrent units (GRU)~\cite{roy2019rf} networks, and transformer~\cite{shen2021asilomar}, are particularly good at encoding discriminative data features in the latent space.

Despite their excellent classification performance~\cite{jian2020deep,shen2021radio,roy2019rf}, deep neural networks are prone to  adversarial attacks, also known as evasion attacks~\cite{goodfellow2014explaining,madry2018towards,bao2021threat}. It has been once and again demonstrated that a well-trained deep learning network can be deluded into incorrect prediction simply by adding maliciously constructed small perturbations to its input. The maliciously perturbed inputs, known as adversarial examples, can be made very subtle and imperceptible to the user or even automated inspection tools. A popularly cited  example is the addition of visually undetectable noise in the picture of a panda to trick a deep neural network into misclassifying it as a 
gibbon~\cite{goodfellow2014explaining}.
Adversarial examples have been widely studied and have evolved rapidly in computer vision tasks but they have recently been extended to the wireless domain, e.g. the modulation classification attack reported in~\cite{kim2021channel, flowers2019evaluating}.

To the best knowledge of the authors, there are only two investigative studies~\cite{bao2021threat,restuccia2020generalized} of adversarial examples on deep learning-based RFFI. In~\cite{bao2021threat}, the authors have examined the effects of adversarial examples on CNNs, but not other deep learning models used in RFFI. In~\cite{restuccia2020generalized}, the evaluation of adversarial examples on a deep learning model is restricted to only I/Q samples. Deep learning models that are trained using other forms of signal representation have not been evaluated. 

This paper aims to investigate the vulnerability of deep learning models in RFFI systems. Two attack methods were investigated, namely fast gradient sign method (FGSM)~\cite{goodfellow2014explaining} and projected gradient descent (PGD)~\cite{madry2018towards}. 
We built a long-range (LoRa) testbed and collected real datasets.
Three deep learning models, namely CNN, LSTM, and GRU were built.
Our main contributions are listed as follows.
\begin{itemize}
\item Adversarial examples produced by FGSM or PGD can severely degrade the identification performance. In a CNN-based RFFI system, 92.7\% and 94.6\% of devices were shown to be successfully misclassified by FGSM and PGD attacks, respectively.  
\item We experimentally corroborated that low quality signals are more vulnerable to evasion attacks. Under $\emph{SNR}=50$~dB, 42.2\% of the devices in RFFI using GRU were misclassified by PGD. When the SNR decreased to 20~dB, 91.3\% of the devices were misclassified.
\item We showed that a targeted attack by PGD is feasible. Superimposing elaborately generated perturbations onto the inputs can mislead the deep learning model to output a specific label set by the adversary. In the experiment of a targeted attack, up to 93.2\% of packets are mistaken to be transmitted from the wrong target device.
\end{itemize}

The rest of the paper is organized as follows. Section~\ref{sec:RFFI} introduces the RFFI process. Adversarial examples for RFFI are presented in Section~\ref{sec:AAinRFFI}. Section~\ref{sec:experi_eva} introduces the design details of the RFFI system and evaluates the effectiveness of adversarial examples on different deep learning models in the RFFI system. Section~\ref{sec:conclusion} concludes the paper.

\section{Radio Frequency Fingerprint Identification}\label{sec:RFFI}
\begin{figure}[!t]
\centering
\includegraphics[width=3.4in]{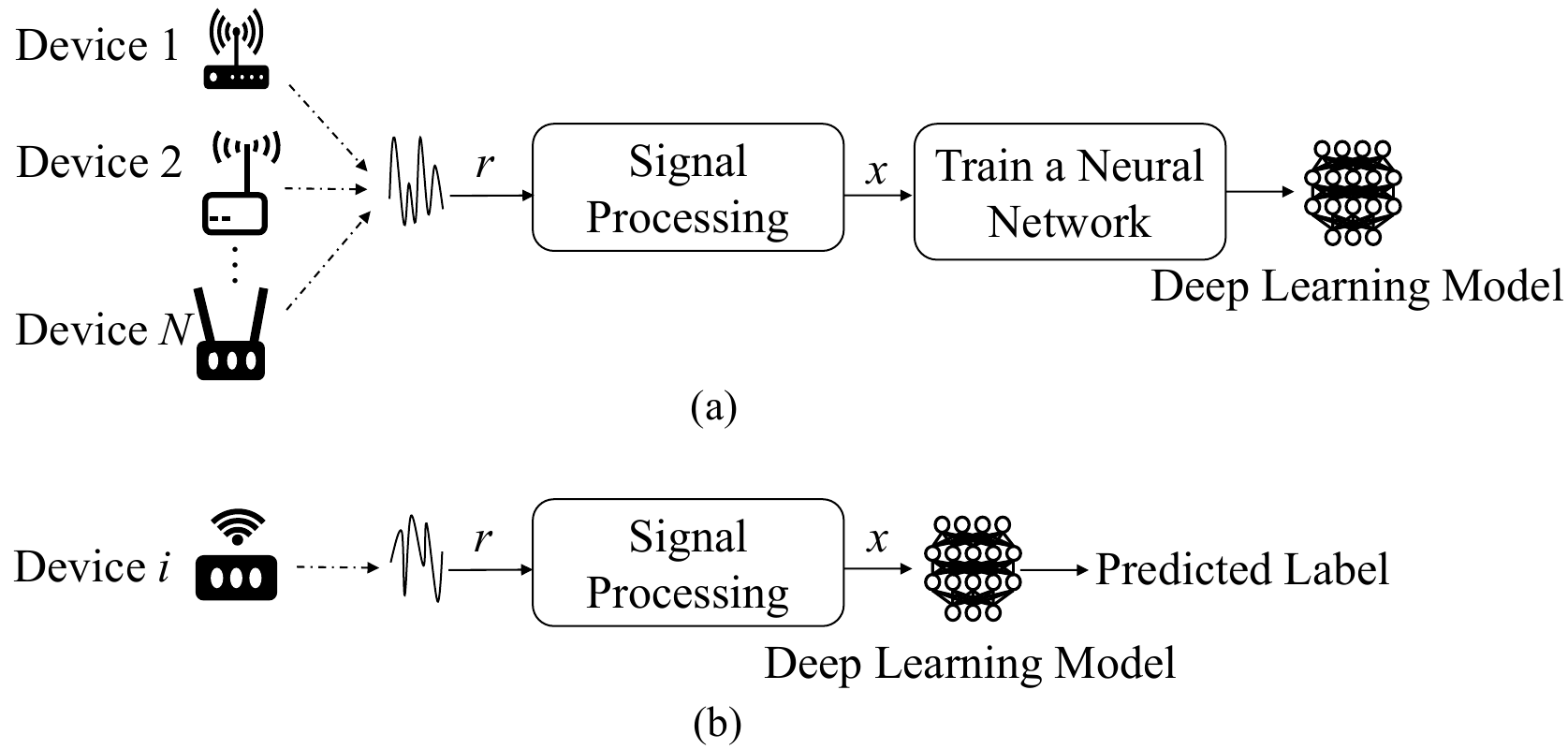}
\caption{RFFI overview (a) Training stage. (b) Inference stage.}
\label{fig:RFFIsys}
\end{figure}

A deep learning-based RFFI system is illustrated in Fig.~\ref{fig:RFFIsys}. 
In the training stage, the deep learning model is trained with RF signals transmitted from $N$ known legitimate devices. In the inference stage, the trained model is used to predict which device has transmitted the signals it receives. 

\subsection{Training}

\subsubsection{Signal Processing}

The received signals, denoted as $r$, need to be processed according to RFFI requirements. The pre-processing usually includes packet detection, synchronization, normalization, CFO estimation and compensation. Detailed algorithms can be found in~\cite{shen2021radio}. 

The captured IQ samples are complex numbers, which cannot be processed by most neural networks. They are transformed to a specific signal representation $x$ before being sent into the neural network for training. The signal representations used in previous work include FFT results~\cite{al2020exposing}, spectrogram~\cite{towardscalable}, error signal~\cite{merchant2018deep}, etc.

\subsubsection{Training} 
A deep learning-based RFFI system needs to be adequately trained before deployment. In the training process, the parameters $\theta$ of the deep learning model $f(\cdot;\theta)$ are continuously updated with the following objective: 
\begin{equation}\label{eqn:train}
\theta=\mathop{\argmin}_{\theta} \sum_{(x, l^{true}(x))} J(f(x;\theta), l^{true}(x)),
\end{equation}
where $x$ is the input to the deep learning model and $f(x;\theta)$ is its corresponding output. $l^{true}(x)$ denotes the ground truth label of $x$. $J(\cdot)$ is the loss function. 

\subsection{Inference}
The inference stage is illustrated in Fig.~\ref{fig:RFFIsys}(b). 
The received signal is processed by the same signal processing module as in the training stage. The transformed signal is fed into the trained deep learning model $f(\cdot;\theta)$. The output $f(x;\theta)$ is a probability vector over all the labels. The index $k$ of the element with the highest probability is selected as the predicted device label $l^{x}$ of the input $x$. The inference process can be mathematically expressed as
\begin{equation}
l^{x} = \mathop{\argmax}_{k}(f(x;\theta)).
\end{equation}


\section{Evasion Attacks on RFFI}\label{sec:AAinRFFI}

\subsection{Threat Model}
An evasion attack in the form of adversarial examples is mounted in the inference stage of RFFI. This paper considers white-box attacks, which assume an adversary has full knowledge of the victim, including training data, deep learning model, etc~\cite{sadeghi2019physical}. Therefore, the adversary employs the processed signal $x$ and model $f(x;\theta)$ to generate a perturbation $v$.
The adversarial example, $x^\prime$, can be written as
\begin{equation}\label{eqn:nosnr}
x^\prime = x + v,
\end{equation}

\begin{figure}[!t]
\centering
\includegraphics[width =3.4in]{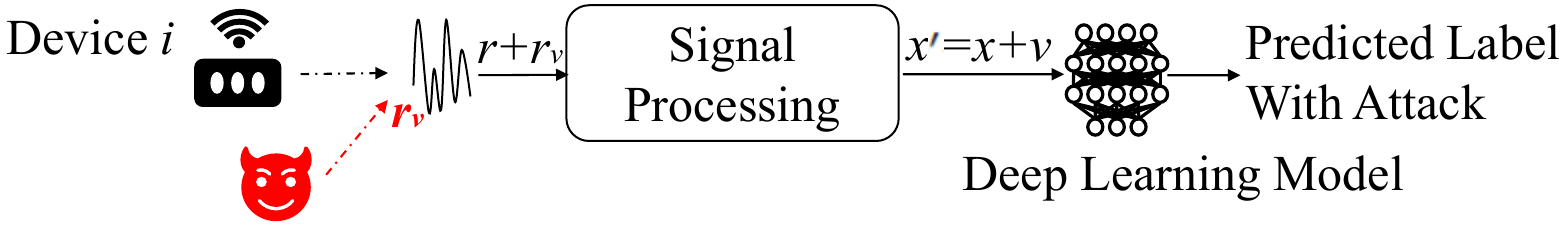}
\caption{Evasion attack on an RFFI system. }
\label{fig:RFFI_with_attack}
\end{figure}

As shown in Fig.~\ref{fig:RFFI_with_attack}, adversarial examples are input to the pre-trained deep learning model to fool the model into making different predictions, given as
\begin{equation}
l^{x'} = \mathop{\argmax}_{k}(f(x';\theta)).
\end{equation}

Evasion attacks can be categorized into non-targeted and targeted attacks based on the adversary’s goal. In the non-targeted attack, the aim is to make the prediction of the adversarial example different from the original label, i.e.,
\begin{equation}
l^{{x}^{\prime}} \ne l^{x}.
\end{equation}

The goal of the targeted attack is to have all adversarial examples predicted to be the target label, i.e.,
\begin{equation}\label{eqn:TAadv}
\begin{aligned}
l^{t} == l^{x'},
\end{aligned}
\end{equation}
where $l^{t}$ is the target label set by the adversary. `$==$' denotes verifying whether the equation on the left equals the right one.

\subsection{Metric}

Perturbation to Signal Ratio (PSR) is the ratio of the power of the perturbation, $P_v$ to the power of the original signal, $P_x$. PSR can be expressed mathematically as:
\begin{equation}\label{eqn:psr}
PSR=10 \log_{10} (\frac{P_{v}}{P_{x}}).
\end{equation}

Success Rate (SR) is defined as the probability of success for an adversarial attack. For the non-targeted attack, it can be evaluated by: 
\begin{equation}\label{eqn:foolingrate}
 SR = \frac{\sum_{i=1}^{N} \zeta \left ( l^{{x}_{i}^{\prime}} \ne l^{{x}_{i}} \right ) }{N},
\end{equation}
where $N$ is the total number of test samples, 
${\zeta}(\cdot)$ is an indicator function that returns one if its argument is true (attack successfully) and zero otherwise. 
For targeted attacks with the target label $l^{t}$, SR can be evaluated by:
\begin{equation}\label{eqn:foolingratetargeted}
 SR = \frac{\sum_{i=1}^{N} \zeta \left ( l^{{x}_{i}^{\prime}} == l^{t}  \right )}{N},
\end{equation}

\subsection{Adversarial Examples Generation Methods}
Among the white-box adversarial examples generation methods, Fast Gradient Sign Method (FGSM) and its iterative version, Projected Gradient Descend (PGD) method, are most popular and are bases for other improved evasion attacks. 

\subsubsection{Fast Gradient Sign Method (FGSM)}
FGSM is a non-iterative approach to produce perturbations that can reliably cause a wide variety of deep neural networks to malfunction with a single projected gradient step (one-step). It can be designed as targeted and non-targeted attacks~\cite{goodfellow2014explaining}. FGSM generates perturbations by calculating the gradient of the loss function relative to the neural network input. Only the gradient sign is used by FGSM, not the whole gradient information. 

Specifically, the adversarial examples for a non-targeted attack are constructed by
\begin{equation}\label{eqn:fgsmnon}
   x^{\prime}=x+\varepsilon \cdot \operatorname{sign}\left(\nabla_{x} J(f(x;\theta), l^{x})\right),
\end{equation}
and adversarial examples for targeted attacks are generated by
\begin{equation}\label{eqn:fgsmtar}
x^{\prime}=x-\varepsilon \cdot \operatorname{sign}\left(\nabla_{x} J(f(x;\theta), l^{t})\right),
\end{equation}
where $\nabla_x$ indicates the gradient of the model for an original sample $x$ with the label $l$, $\varepsilon$ denotes a control parameter for the amplitude of perturbation, $\operatorname{sign}(\cdot)$ indicates the gradient direction. $\operatorname{sign}(\cdot)$ returns 1 (-1) if the value of the gradient direction is greater (smaller) than 0.

\subsubsection{Projected Gradient Descent (PGD)}
PGD is an iterative method with random initialization to produce adversarial examples. It iteratively applies FGSM and projects the perturbed sample back to the norm multiple times with step size in the scale of the total perturbation bound. PGD is more powerful than FGSM because it mitigates the underfitting adversarial example due to the linear approximation of the decision boundary around the data point.
The procedure of PGD is given in Algorithm~\ref{algorithm:pgd}, where $\left \| \cdot \right \| _p$ is the norm of the perturbation (such as $L_2$ and $L_\infty$). $\operatorname{Clip}_{x, \alpha}(\cdot)$ denotes clipping the argument to the range $[x-\alpha, x+\alpha]$.



\begin{algorithm}[!t]
\DontPrintSemicolon
  \KwInput{signal ${x}$, label $l$ or $l^{t}$, loss function ${J(\cdot)}$, size of the perturbation $\varepsilon$, the size on each iteration step $\alpha$, number of iteration $I$ }
  \KwOutput{An adversarial example $ x_{I}^{\prime}$ with $\left\|x_{I}^{\prime}-x\right\|_{\text {p }} \leq \varepsilon$}
Initialize $x_0^{\prime} = x$, $i=1$

\For{$i$ in range $I$}{
\If{Non-targeted attack}
{$x_{i}^{\prime}=\operatorname{Clip}_{x, \alpha}\left\{x_{i-1}^{\prime}+\alpha \cdot \operatorname{sign}\left(\nabla_{x} J(f(x;\theta), l^{x})\right)\right\}$
}

\ElseIf{Targeted attack}
{$x_{i}^{\prime}=\operatorname{Clip}_{x, \alpha}\left\{x_{i-1}^{\prime}-\alpha \cdot \operatorname{sign}\left(\nabla_{x} J(f(x;\theta), l^{t})\right)\right\}$
}
}

\Return $ {x_I^{\prime}}$
\caption{PGD Algorithm}\label{algorithm:pgd}
\end{algorithm}

\section{Experimental Evaluation}\label{sec:experi_eva}

\subsection{Setup}
\subsubsection{Datasets}
LoRa is used as a case study in this paper. 
Eight preambles at the start of each LoRa packet are captured for RFFI. The IQ samples are then converted into channel-independent spectrograms, which are in the time-frequency domain and robust to the channel variation. We refer interested readers to~\cite{towardscalable} for detailed derivation. 

\begin{figure}[!t]
\centering
\subfloat[]{\includegraphics[width=1.5in]{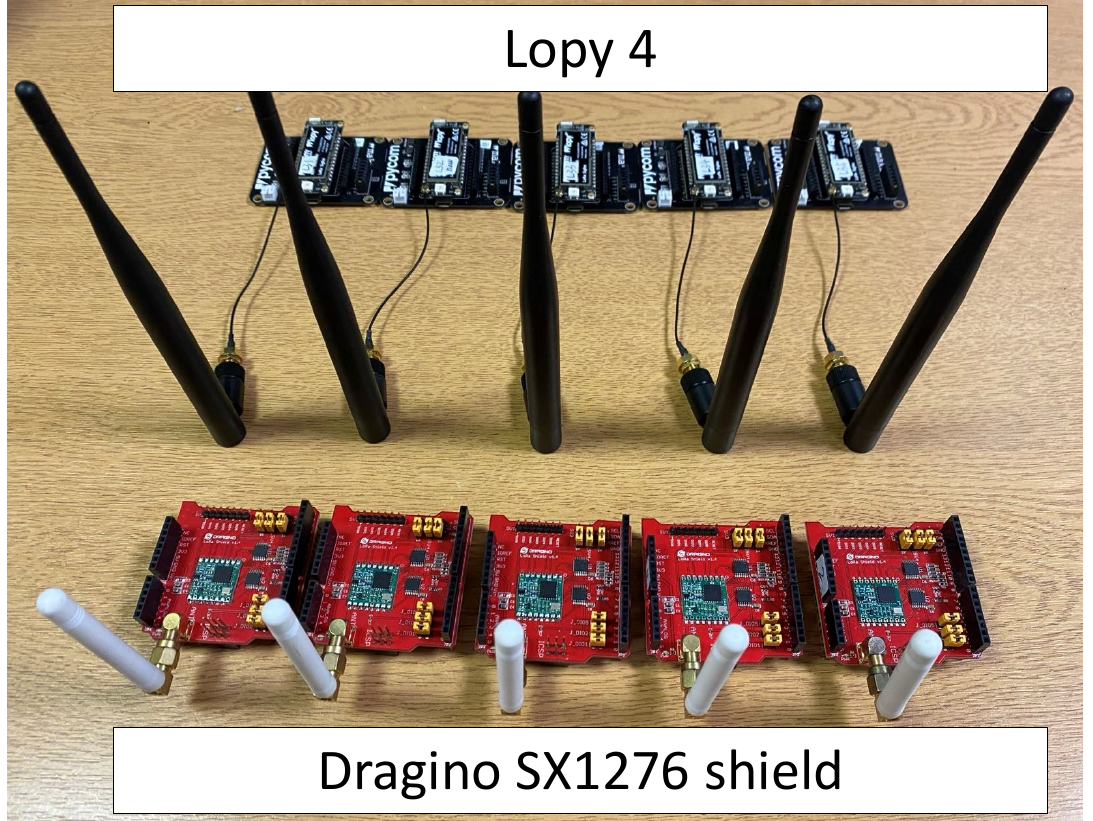}\label{fig:lora}}
\subfloat[]{\includegraphics[width=1.5in]{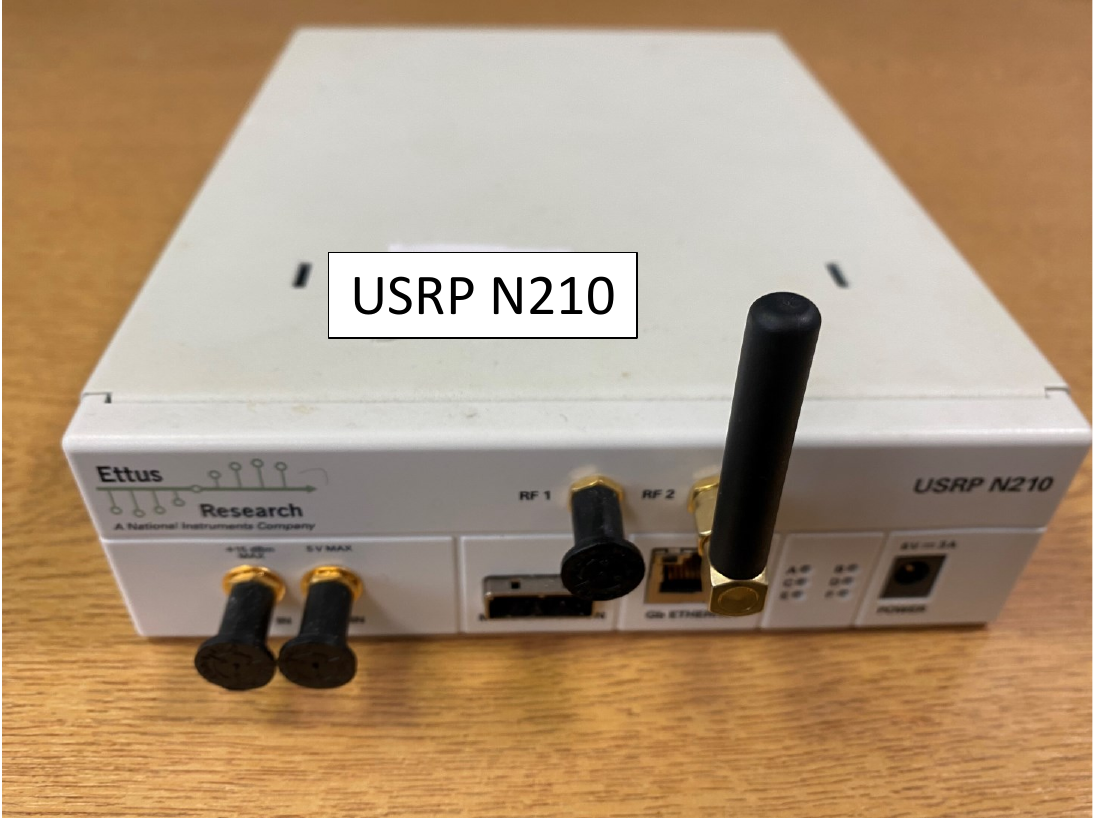}\label{fig:usrp}}

\centering
\caption{Experimental devices. (a) Transmitter: LoRa. (b) Receiver: USRP. }
\label{fig:equip}
\end{figure}

As shown in Fig.~\ref{fig:equip}(a), ten LoRa devices, consisting of five LoPy4 and five Dragino SX1276 shields, are used. To distinguish these ten devices, we label them as Device 1, Device 2 ... Device 10, respectively.
These devices are configured with bandwidth $B = 125$~kHz and carrier frequency $f_c = 868.1$~MHz. The receiver end is a USRP N210 software-defined radio (SDR) platform, as shown in Fig.~\ref{fig:equip}(b), whose sampling rate $f_s = 1$~MHz. The Communications Toolbox Support Package for USRP Radio of MATLAB R2021b is used for accessing data from USRP. The packets are collected in an office environment and the SNR is about 70~dB. The datasets contain 1000 packets from each device. For each device, 90\% and 10\% of the packets are used for training and testing, respectively. 
   
\subsubsection{Models} 
\begin{figure}[t]
\centering
\includegraphics[width=3.2in]{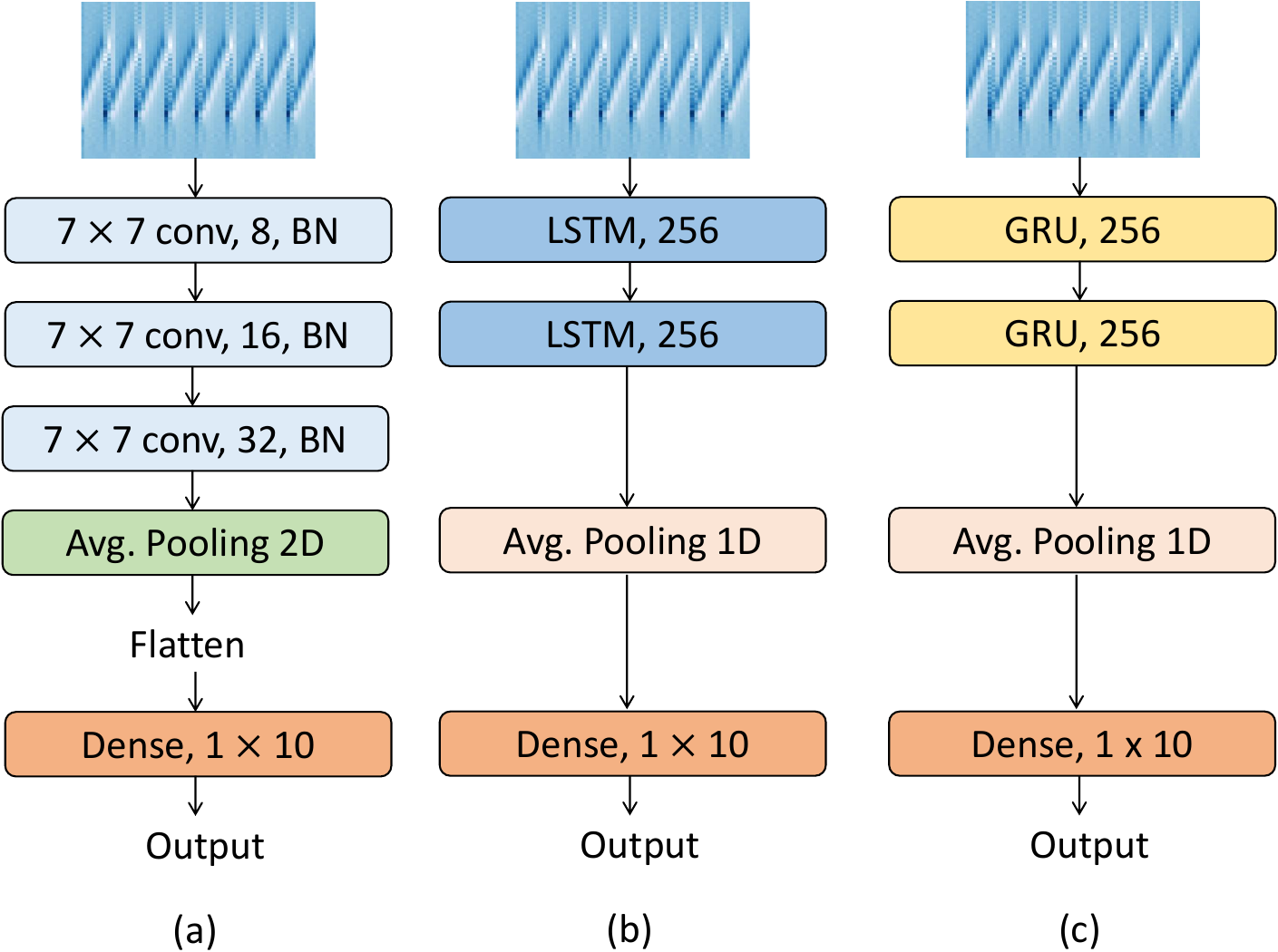}\label{fig:CNN}
\caption{Three deep learning network architectures: (a) CNN. (b) LSTM network. (c) GRU network.}
\label{fig:2}
\end{figure}

We apply adversarial attacks to three popular deep learning models, namely CNN, LSTM, and GRU, whose architectures are shown in Fig.~\ref{fig:2}.
\begin{itemize}
\item CNN (Fig.~\ref{fig:2}(a)): it consists of one convolutional layer of 8 $7 \times 7$ kernels, followed by one convolutional layer of 16 $7 \times 7$ kernels and one convolutional layer of 32 $7 \times 7$ kernels, one 2D average pooling layer, and one dense layer activated by the softmax function.
\item LSTM (Fig.~\ref{fig:2}(b)): it consists of two 256-unit LSTM layers, one global 1D average pooling layer, and one softmax activated dense layer. 
\item GRU (Fig.~\ref{fig:2}(c)): it  is similar to the LSTM (Fig.~\ref{fig:2}(b)), but the LSTM layers are replaced by the GRU layers.
\end{itemize}

We use a prefix-suffix notation ``\textit{A}-\textit{B}" to denote an adversarial example generation method \textit{A} mounted on a deep learning model \textit{B}. For example, ``FGSM-CNN'' refers to the attack on a CNN-based RFFI system using the adversarial examples generated by FGSM. 
   
\subsubsection{Training Configuration} 
The deep learning training was carried out on a PC equipped with a GPU of NVIDIA GeForce GTX 1660Ti. The neural networks are implemented with the TensorFlow library. FGSM and PGD are implemented based on CleverHans.v3.1.0~\cite{papernot2018cleverhans}, which is a software library to benchmark machine learning systems' vulnerability to adversarial examples. The standardized reference implementations of attacks are provided in this library.
  
\subsection{Comparison Between Clean Sample and Adversarial Example} 
\begin{figure}[!t]
\centering
\subfloat[]{\includegraphics[width=1.18in]{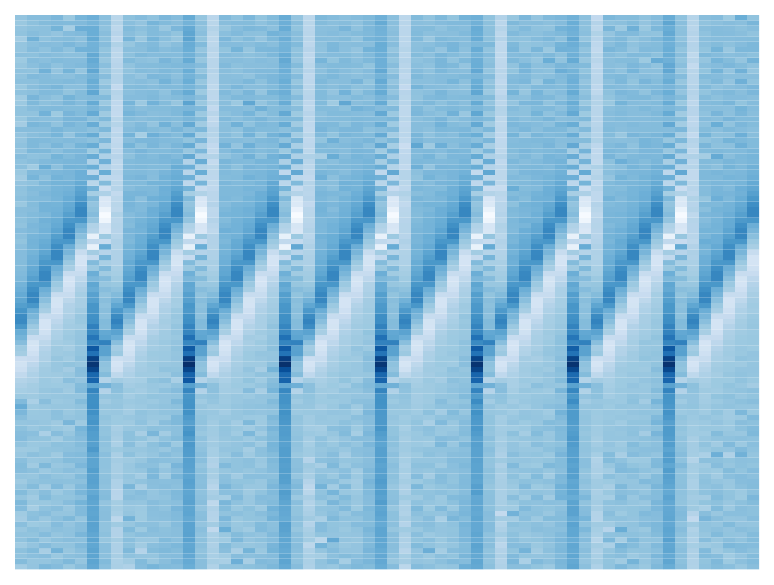}\label{fig:spec_wo_aa}}
\subfloat[]{\includegraphics[width=1.18in]{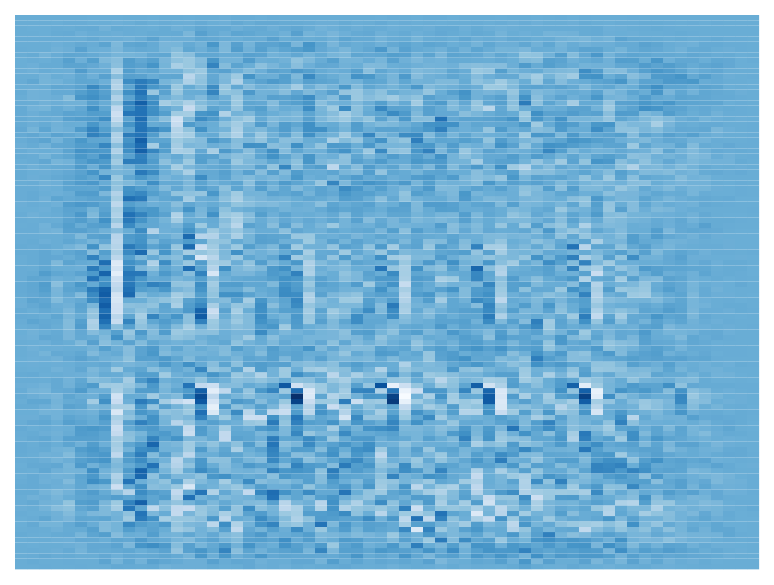}\label{fig:pgdv}}
\subfloat[]{\includegraphics[width=1.18in]{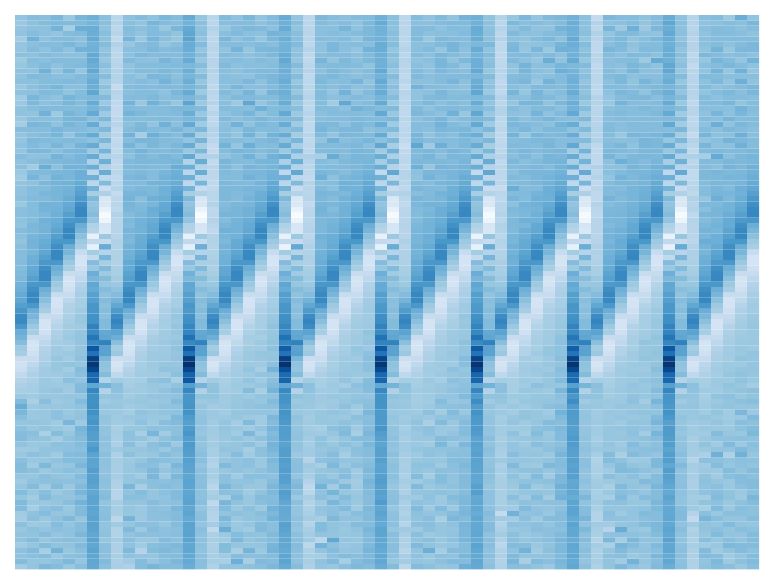}\label{fig:spec_w_aa}}
\caption{(a) Channel independent spectrogram of clean sample, $x$.  (b) Perturbation calculated by PGD, $v$. (c) Adversarial example, $x^{\prime}$.}
\label{fig:3}
\end{figure}

A channel-independent spectrogram of a clean sample, its corresponding perturbation and adversarial example are illustrated in Fig.~\ref{fig:3}.
\begin{itemize}
	\item Fig.~\ref{fig:3}(a) is the channel-independent spectrogram of one sample in the original datasets. We can observe eight repeating upchirps from Fig.~\ref{fig:3}(a), and these upchirps contain unique features of the device. 
	\item Fig.~\ref{fig:3}(b) is the perturbation calculated by PGD based on the spectrogram shown in Fig.~\ref{fig:3}(a).
	\item Fig.~\ref{fig:3}(c) is the corresponding adversarial example. 
\end{itemize}
   
   Visually we can barely distinguish the spectrogram of the clean sample and adversarial example in Fig.~\ref{fig:3}. Therefore, an objective metric is required to quantify the distortion of the sample and the effectiveness of the attack.
   In the adversarial example generation algorithms, the maximum perturbation of the sample is controlled by $\varepsilon$ ((\ref{eqn:fgsmnon}), (\ref{eqn:fgsmtar}) and Algorithm~\ref{algorithm:pgd}). In this paper, we use PSR to assess the power of perturbation signal, and then we measure the success rate SR of different algorithms. PSR and SR enable an objective quantitative evaluation of the effectiveness of using different adversarial attack methods to spoof a wireless communication receiver.

\subsection {Non-targeted Attacks With Different Perturbations} \label{sec:diffPSR}
To demonstrate the effectiveness of different adversarial examples, we compare them with additive white Gaussian noise (AWGN) which is a common natural disturbance in RFFI systems. We show the confusion matrices for different perturbations and mounted FGSM and PGD on RFFI to compare their variations in SR against PSR.

\begin{figure}[!t]
\centering
\subfloat[]{\includegraphics[width=1.7in]{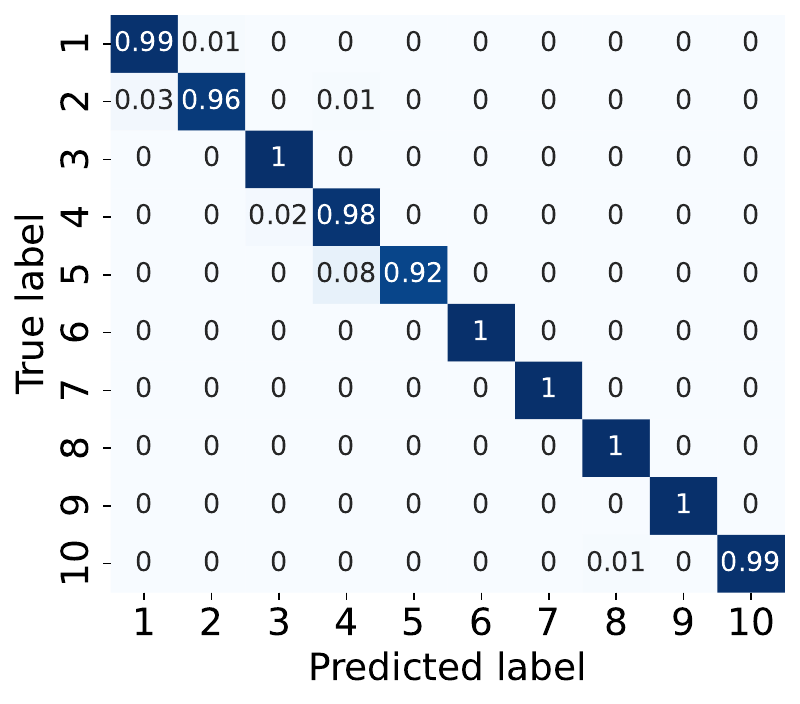}\label{fig:12}}
\hspace{0.1cm}
\subfloat[]{\includegraphics[width=1.7in]{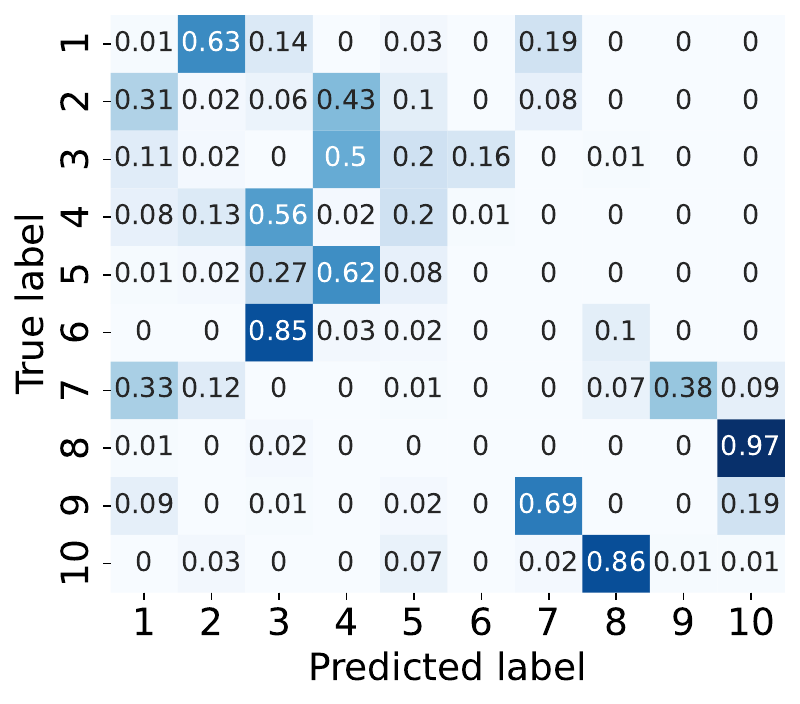}\label{fig:22}}
\\
\subfloat[]{\includegraphics[width=1.7in]{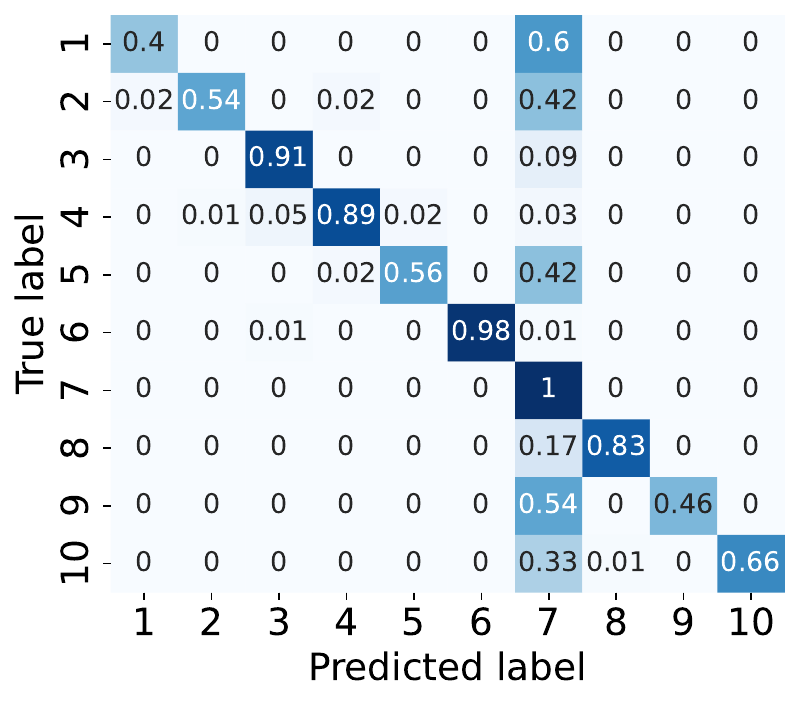}\label{fig:13}}
\hspace{0.1cm}
\subfloat[]{\includegraphics[width=1.7in]{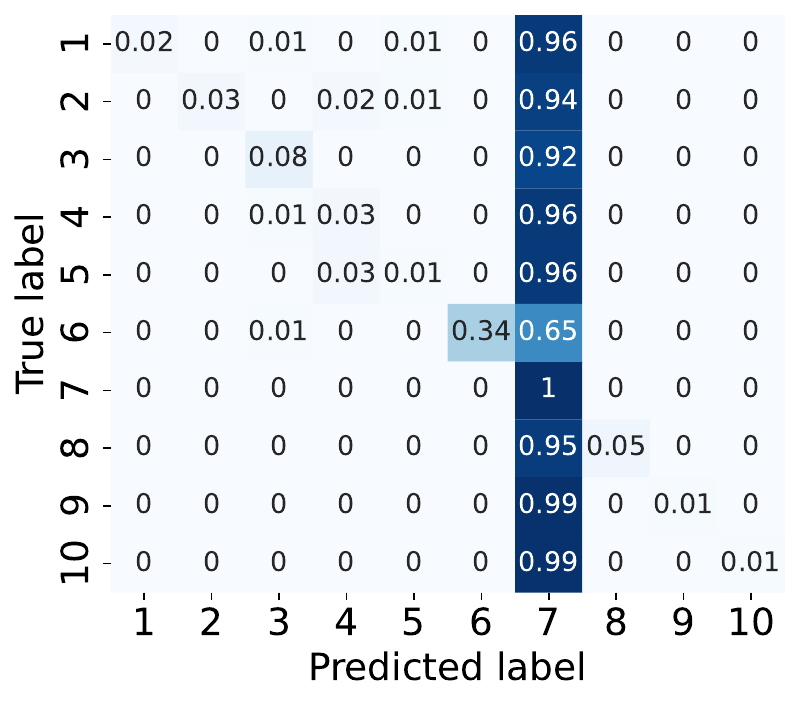}\label{fig:14}}
\caption{Confusion matrix. (a) RFFI without adversarial attack. (b) CNN-PGD with non-targeted attacks. $\emph{PSR}=-35$~dB. (c) CNN-PGD with targeted attacks. $\emph{PSR}=-30$~dB.  (d) CNN-PGD with targeted attacks. $\emph{PSR}=-25$~dB. }
\label{fig:15}
\end{figure}

Fig.~\ref{fig:15}(a) shows the confusion matrix of the CNN-based RFFI without attack and the classification accuracy reaches 98.4\%. Fig.~\ref{fig:15}(b) shows the confusion matrix of non-targeted attack produced by PGD when $\emph{PSR}=-35$~dB and the overall accuracy reduces to $1.4\%$. As it is a non-targeted attack, the misclassification results are dispersed across all labels.

\begin{figure}[t]
\centering
\includegraphics[width =3in]{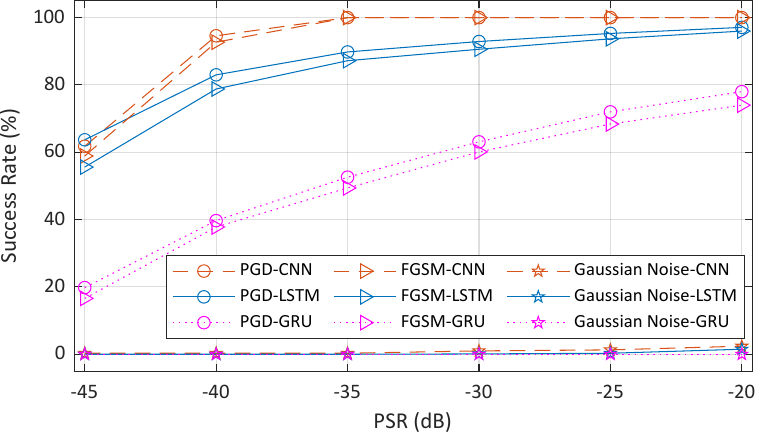}
\caption{SR of non-targeted attacks under different PSRs.}
\label{fig:7}
\end{figure}

Fig.~\ref{fig:7} shows the SR of non-targeted attacks when PSR changes from $-45$~dB to $-20$~dB.
Deep learning-based RFFI systems suffer more severe misclassifications when PSR is higher. Specifically, when the PSR is $-40$~dB, The SRs of CNN-PGD RFFI and CNN-FGSM RFFI reach 94.6\% and 92.7\%, respectively. 
The SRs of all the adversarial attacks continue to increase with increasing PSR. Also, the iterative PGD is found to be more effective than the one-step FGSM. 

Fig.~\ref{fig:7} shows that the same power of AWGN, when injected into the input signal, seldom has any impact on the performance of these RFFI classifiers.
Apparently, while RFFI classifiers are robust against AWGN, they are vulnerable to adversarial examples as evinced by their significantly compromised prediction accuracy under these attacks.

\subsection{Non-targeted Attacks Under Different SNRs} \label{sec:diffSNR}
The attacks in the previous sections assume good quality datasets are acquired under a high SNR. However, in a real RFFI system, AWGN will reduce the SNR. Its effect on the quality of the datasets cannot be ignored. Taking AWGN into consideration, $x^{\prime}$ is expressed in (\ref{eqn:insnr}). 
\begin{equation}\label{eqn:insnr}
x^{\prime} = x + n + v,
\end{equation}
where $n$ is the AWGN.

The former results were obtained under $\emph{SNR}=70$~dB, which means the quality of the datasets is high. In this section, we emulate different SNRs by adding artificial AWGN into the original clean datasets. 

\begin{figure}[t]
\centering
\includegraphics[width =3in]{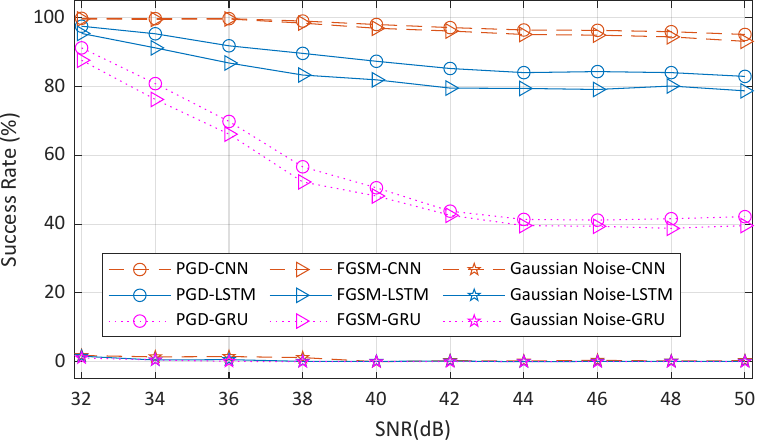}
\caption{SR of non-targeted attacks at different SNRs. $\emph{PSR}=-40$ dB.}
\label{fig:11}
\end{figure}

Fig.~\ref{fig:11} presents the SRs of different models and adversarial attack methods against SNRs. It can be observed that the effectiveness of all evasion attacks decreases at different rates with increasing SNR. Taking FGSM-GRU as an example, the SR is about $87.7\%$ when $\emph{SNR}=32$~dB but it reduces to about $39.6\%$ when $\emph{SNR}=50$~dB.
The adversarial examples are much more likely to succeed when the SNR is low, regardless of the deep learning model. It implies that an active adversary can monitor the received signal strength and channel characteristics and seize a good opportunity to mount the attack to increase the attack success rate.

\subsection{Targeted Attacks} \label{sec:pgd-target}
PGD is utilized to implement targeted attacks. We arbitrarily select Device 7 as the target.
Fig.~\ref{fig:15}(c) shows the confusion matrix of PGD-CNN with the targeted attack at $\emph{PSR}=-30$~dB, where 36.1\% of the devices are misclassified as Device 7. 
When the PSR increases to $-25$~dB, 93.2\% of the devices are misclassified as Device 7, as shown in Fig.~\ref{fig:15}(d). A successful targeted attack is hence demonstrated.
The effectiveness of the targeted attack can be further enhanced by higher PSR to cause more devices to be misclassified as Device 7.

\begin{figure*}[!t]
\centering
\subfloat[]{\includegraphics[width=2.3in]{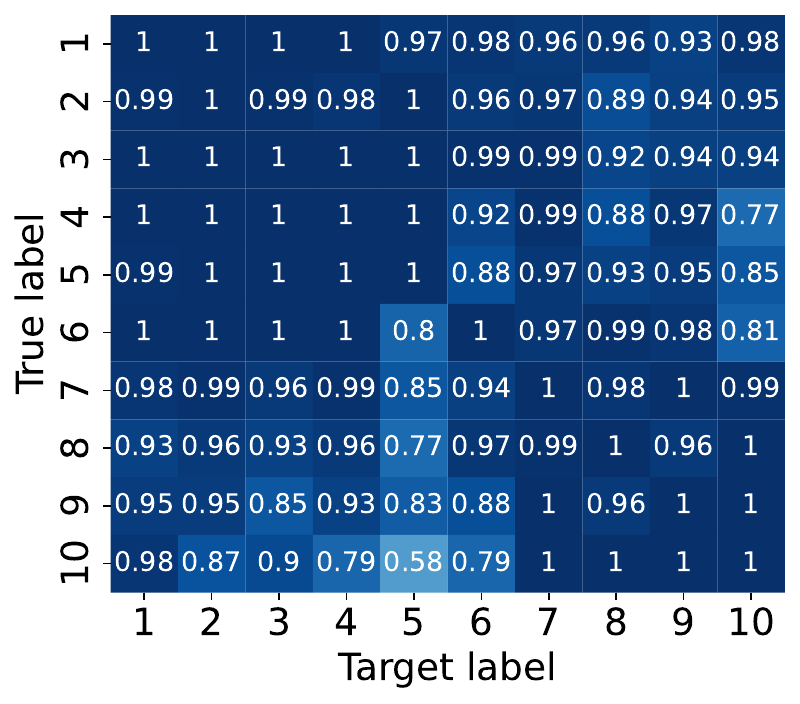}\label{fig:16}}
\subfloat[]{\includegraphics[width=2.3in]{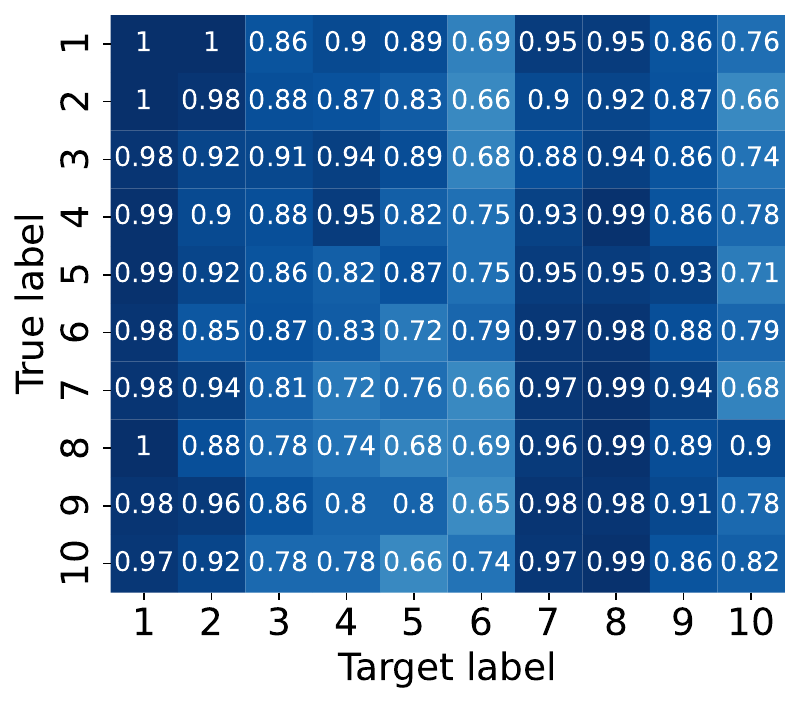}\label{fig:17}}
\subfloat[]{\includegraphics[width=2.3in]{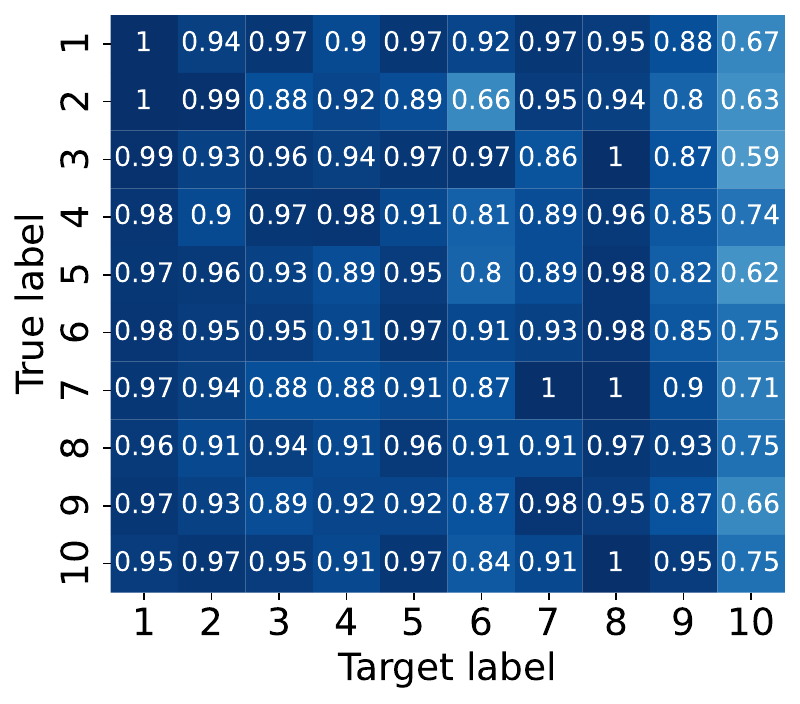}\label{fig:18}}
\centering
\caption{Targeted attacks by PGD on (a) CNN, (b) LSTM, and (c) GRU-based RFFI, with $\emph{PSR}=-5$ dB.}
\label{fig:99}
\end{figure*}
To investigate the performance of targeted attacks against different targets, we set each class as the target one by one. The results are shown in Fig.~\ref{fig:99}. Unlike a confusion matrix, each matrix element in Fig.~\ref{fig:99} represents the percentage of devices identified as the selected target. For example, 80\% packets from Device 6 are identified as originating from Device 5 in Fig.~\ref{fig:99}(a) when the selected target is Device 5. Apparently, PGD can cause the victim model to misclassify most devices as the selected target indiscriminately, which indicates the effectiveness of PGD for targeted attacks on deep learning-based RFFI. 

\section{Conclusion}\label{sec:conclusion}
In this paper, we presented white-box evasion attacks on deep learning-based RFFI. 
Specifically, we studied non-targeted and targeted attacks using two adversarial example generation methods, namely FGSM and PGD. 
A LoRa-RFFI testbed was built and real datasets were collected.
The attacks were applied to CNN, LSTM and GRU-based RFFI systems.
The results reveal that deep learning-based RFFI systems are vulnerable to adversarial attacks. 
Compared with Gaussian noise, the carefully crafted perturbation can completely cripple the RFFI.
The iterative method PGD was demonstrated to be more effective than the one-step method FGSM, regardless of SNR. 
In short, the successes of both targeted and non-targeted attacks signify a real threat to the use of modern RFFI for physical layer security.
In a white-box attack, the adversary has all the knowledge of the victim and is able to calculate the accurate gradient to result in a significant security problem.
In our future work, we will investigate more practical black-box or grey-box attacks where the adversary has no knowledge about the victim's deep learning model. We will also develop lightweight countermeasures against these successful evasion attacks.


\section*{Acknowledgement}
The work was in part supported by the UK EPSRC New Investigator Award under grant ID EP/V027697/1 and in part by National Key Research, Development Program of China under grant ID 2020YFE0200600 and in part by the National Natural Science Foundation of China under grant ID 62171121.
The work of J.~Ma was also supported by the China Scholarship Council CSC under Grant ID 202006470007. The work of C. H. Chang was supported by the Ministry of Education, Singapore, under its AcRF Tier 2 Award No MOET2EP50220-0003.

\bibliographystyle{IEEEtran}
\bibliography{IEEEabrv,mybibfile}

\end{document}